\documentclass{elsart5p}

\usepackage{graphicx}
\usepackage{graphics}
\usepackage{amssymb}
\usepackage{bm}
\usepackage{braket}
\usepackage{float}
\usepackage{color}
\usepackage{amsmath}

\journal{Physica B: Condensed Matter}

\begin{document}

\begin{frontmatter}

\title{Impurities or a neutral Fermi surface? A further examination of the low-energy ac optical conductivity of SmB$_6$}

\author[JHUIQIM]{N. J. Laurita
\corauthref{mycorrespondingauthor}}
\corauth[mycorrespondingauthor]{Corresponding author}
\ead{Laurita.Nicholas@gmail.com}
\author[JHUIQIM] {C. M. Morris}
\author[JHUIQIM]{S. M. Koohpayeh}
\author[JHUIQIM,JHUChem]{W. A. Phelan}
\author[JHUIQIM,JHUChem,JHUMat]{T. M. McQueen}
\author[JHUIQIM]{N. P. Armitage}

\address[JHUIQIM]{The Institute For Quantum Matter, Department of Physics and Astronomy, The Johns Hopkins University, Baltimore, Maryland, 21218, USA}
\address[JHUChem]{Department of Chemistry, The Johns Hopkins University, Baltimore, Maryland, 21218, USA}
\address[JHUMat]{Department of Materials Science and Engineering, The Johns Hopkins University, Baltimore, Maryland, 21218, USA}

\begin{abstract}

Recent experiments have uncovered evidence of low energy excitations in the bulk of SmB$_6$ that are perhaps associated with unconventional quasiparticles, bringing into question whether this Kondo ``insulator" is truly insulating in the bulk.  Recently, we demonstrated that SmB$_6$ possesses significant in-gap bulk ac conduction far in excess of typical disordered semiconductors. Whether such conduction is an intrinsic feature of SmB$_6$, suggesting the formation of an exotic state, or residual conduction from impurities continues to be a topic of debate.  Here, we further examine the origin of the ac optical conductivity of SmB$_6$ in light of recent experimental and theoretical developments.  The optical conductivity of SmB$_6$ is shown to possess distinct regimes of either dominant free carrier or $localized$ response contributions.  The free carrier response is found to be in good qualitative agreement with previous literature, although quantitative differences are revealed and discussed.  The localized response, which dominates at the lowest temperatures, is analyzed in the context of models of either in-gap impurity states or an exotic neutral Fermi surface. The charge density or effective mass of this low temperature in-gap conductivity is extracted through a conductivity sum rule analysis and found to be in general alignment with both models in the appropriate limits.  Our results shed further light on the nature of the in-gap states of this remarkable material.
\end{abstract}


\end{frontmatter}

\section{Introduction}

Gapped systems are hallmarked by an exponential divergence of their resistivity with reducing temperature.  Deviations from this behavior are indicative of additional conduction mechanisms, whether they be extrinsic, perhaps stemming from impurities or disorder, or intrinsic, for instance resulting from the formation of topological surface states.  In the late 1960's, resistivity measurements \cite{Menth1969,Nickerson1971} performed on the mixed-valent \cite{Varma1976} semiconductor SmB$_6$ revealed two prominent features: First, the onset of such an exponential dependence at T $\approx$ 50K, signifying a crossover from metallic to insulating behavior.  Today this crossover is known to stem from the hybridization of localized 4$f$ electrons near the Fermi level and itinerant 5$d$ electrons \cite{Coqblin1968,Mott1974,Varma1976}, opening a gap in the density of states of order $\Delta_g$ $\approx$ 20 meV.  The second, and more surprising, feature is a plateauing of the resistivity that occurs at temperatures T $<$ 5K, signifying a new dominant conduction mechanism.  Despite intense experimental and theoretical investigation, the physics behind this plateau has remained a mystery for nearly half a century. 

Interest in SmB$_6$ was rekindled in 2010 due to the seminal prediction that SmB$_6$ may be the first known example of a topological Kondo insulator (TKI), a new state of matter in which strong interactions and spin-orbit coupling conspire to open a parity inverted gap with corresponding in-gap topological surface states \cite{Dzero2010}.  In this interpretation, the plateau in the resistivity of SmB$_6$ results from the topological surface states which short out the insulating bulk at low temperatures.  Accordingly, a flurry of theoretical \cite{Dzero2010,Takimoto2011,Dzero2012,Fu2013,Alexandrov2013} and experimental \cite{Kim2013,Kim2014,Wolgast2013,Chen2015,XZhang2013,Yee2013,Li2014,Nuepane2013,Miyazaki2012,Xu2014,Jiang2013,Frant2013,Fuhrman2015} investigations into the physical properties of SmB$_6$ have been reported in recent years, many of which have proclaimed findings consistent, although perhaps not conclusively so, with the TKI prediction.    

The realization of a TKI is certainly an exciting prospect.  It has been proposed that since the gap of SmB$_6$ is interaction driven it may be truly insulating, which has generally not been the case in other classes of non-interacting topological insulators. Therefore, SmB$_6$ may be better suited for technological applications.  Indeed, dc resistivity measurements do suggest an insulating bulk \cite{Kim2013, Kim2014, Wolgast2013, Chen2015} with more recent measurements observing the exponential divergence of the bulk resistivity over an incredible 10 orders of magnitude, including to very low temperatures deep within the plateaued regime \cite{Eo2017}. 

However, the story of SmB$_6$ took an unexpected turn in 2015 when torque magnetometry experiments performed by Tan \textit{et al.} \cite{Tan2015} uncovered what appeared to be quantum oscillations associated with an unconventional 3-dimensional Fermi surface, potentially consisting of exotic neutral quasiparticles.  In fact, evidence of a potentially exotic state within the bulk of SmB$_6$ can be traced back decades.  It has long been known that the Sommerfeld coefficient of the specific heat of SmB$_6$ is unusually large, $\gamma$ $\approx$ 10 mJ/mol*K$^2$ \cite{Flachbart2006,Gabani2001,Phelan2014} - 10 times larger than \textit{metallic} LaB$_6$.  Such large fermionic specific heat has since been shown to be a bulk effect \cite{Wakeham2016}.  Additionally, low-energy ac conductivity experiments of SmB$_6$ have revealed in-gap conduction consistent with a \textit{localized response} with conductivities orders of magnitude larger than the dc value \cite{Molnar1982,Travaglini1984,Jackson1984,Kimura1994,Nanba1993,Gorshunov1999}.  The aforementioned experimental findings raise an important red flag in our understanding of SmB$_6$; is the Kondo insulator SmB$_6$ even...an insulator?

We raised this question in a previous publication based on our low-energy optical spectroscopy experiments \cite{LauritaSmB6}.  In that work we definitively demonstrated that the large in-gap ac conduction of SmB$_6$ is three-dimensional and therefore a bulk effect.  Whether such large in-gap conduction is a hint of an exotic state \cite{Baskaran2015,Onur2017,Chowdhury2017} or simply residual conduction from impurities \cite{Sollie1991,Schlottmann1992,Schlottmann1996,Riseborough2003,Fuhrman2017} continues to be a topic of debate.  However, we argued that there are reasons to believe that such conduction may be intrinsic to SmB$_6$.  For instance, we found that the ac conductivity was highly repeatable, even when comparing samples grown by different techniques - in stark contrast to the exponentially sensitive conduction of impurity band insulators \cite{Helgren2004}.  Furthermore, although the in-gap conductivity displays a power law frequency dependence that resembles impurity band insulators, it is orders of magnitude larger than what is observed in disordered semiconductors and is more on-par with \textit{completely amorphous} alloys \cite{Helgren01a}.  

In this work, we further analyze the optical conductivity of SmB$_6$ in the context of recent theoretical and experimental developments.  As our experimental energy range, 1 - 8 meV, is less than the bulk gap, $\Delta_g$ $\approx$ 20 meV, we directly probe the in-gap conduction of SmB$_6$.  The conductivity is modeled with two contributions - a free carrier response which dominates at high temperatures and a ``localized" response which dominates within the plateaued state.  To be clear, the term localized does not necessarily refer to the charge carriers themselves, but to the response of the current as it is inferred that the dc conductivity goes to zero in the zero frequency limit.  The free carrier response is found to be in good qualitative agreement with previous literature, although quantitative differences are revealed and discussed.  The localized response is analyzed in the context of models of either in-gap impurity states or an exotic neutral Fermi surface. The charge density or effective mass of this low temperature in-gap conductivity is extracted through a conductivity sum rule analysis and found to be in general alignment with both models in the appropriate limits.  Our results shed further light on the nature of the in-gap states of this remarkable material.

\section{Methods}

Highly sensitive time-domain terahertz (TDTS) experiments were performed on single crystal samples of SmB$_6$ grown by both optical floating zone and aluminum flux growth techniques.  In the present paper, we concentrate solely on samples grown by the optical floating zone method.  Such optical transmission experiments are exceptionally difficult to perform on SmB$_6$ due to the material's high index of refraction ($n$ $\approx 25$) and, as demonstrated below, large absorption in the THz range.  Correspondingly, novel methods for performing TDTS experiments on SmB$_6$ were developed.  Details of our methods, which involved mounting SmB$_6$ samples to Al$_2$O$_3$ substrates and polishing to thicknesses of 10's of $\mu$m, can be found in Ref. \cite{LauritaSmB6}.  

TDTS transmission experiments were performed using a home built spectrometer within a temperature range of 1.6K to 300K \cite{Laurita2016} on single crystal SmB$_6$ samples oriented such that the $\hat{c}$ [001] axis was perpendicular to  the plane of the sample surface.  TDTS is a high resolution method for accurately measuring the electromagnetic response of a sample in the experimentally challenging THz range.  In our TDTS experiments, the electric field of a transmitted THz pulse through a SmB$_6$ sample mounted to an Al$_2$O$_3$ substrate was measured as a function of real time. Performing a Fourier transform of the measured time-domain electric field and referencing to the transmitted electric field through an identical substrate allows access to the frequency dependent \textit{complex} transmission spectrum of the sample.  The complex transmission is then given by: 
\begin{equation}
\widetilde{T}(\omega) = \frac{2\widetilde{n}(\widetilde{n}_s + 1)}{(\widetilde{n}+1)(\widetilde{n}+\widetilde{n}_s)}\exp{[\frac{i \omega d}{c}(\widetilde{n}-1)]}
\label{Transeq2}
\end{equation}
where $d$ is the sample thickness, $\omega$ is the frequency, $c$ is the speed of light, $\widetilde{n}$ is the sample's complex index of refraction, $\widetilde{n}_s$ = 3.1 is the Al$_2$O$_3$ substrate's index of refraction, and normal incidence in a vacuum has been assumed.  The complex transmission can then be numerically inverted via a Newton-Raphson algorithm to obtain the sample's complex index of refraction from which the complex optical conductivity can be determined.

\section{Experimental Results}

\begin{figure}
\includegraphics[width=1.0\columnwidth, keepaspectratio]{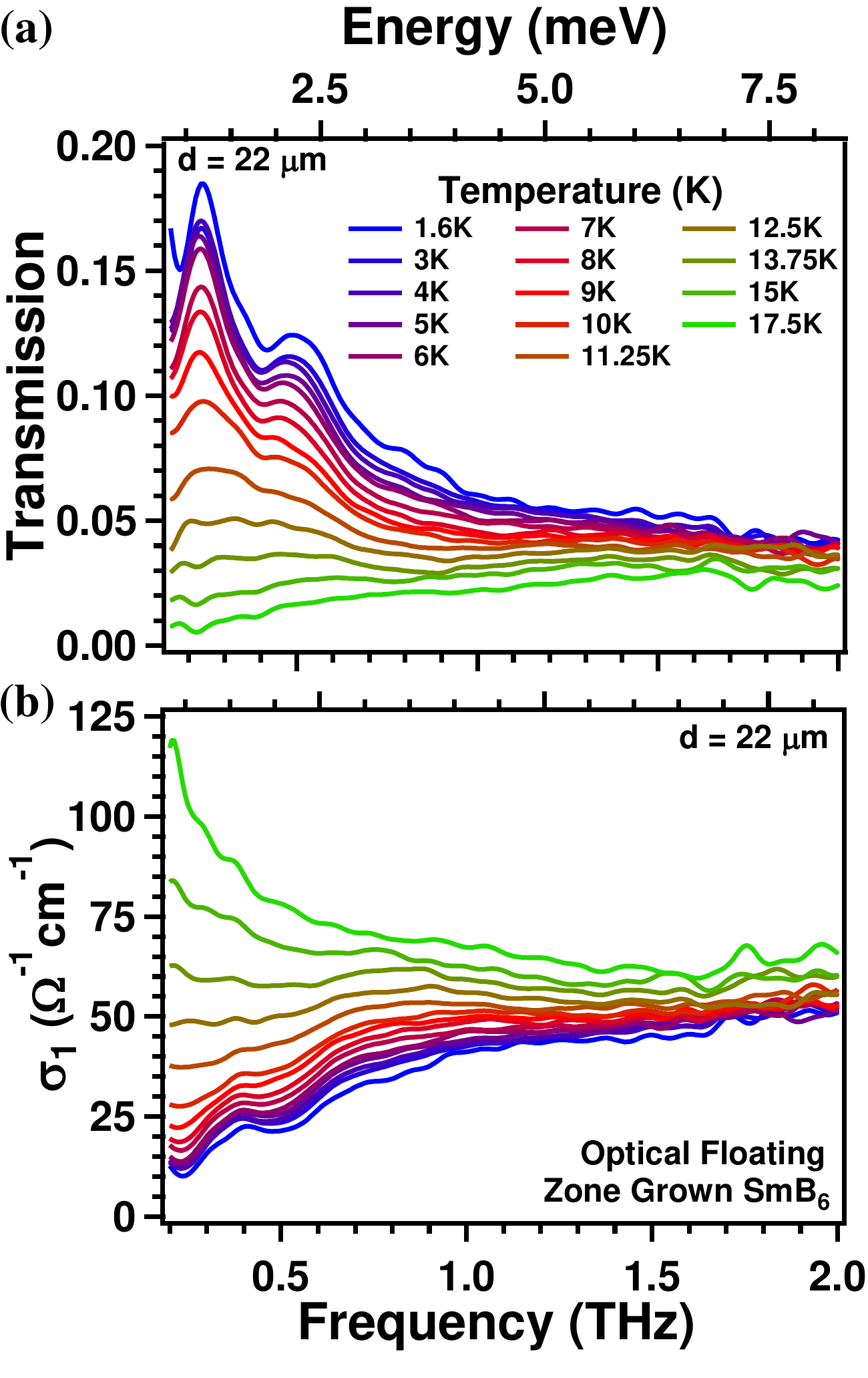}
\caption{(a,b) Magnitude of the complex transmission, as defined in Eq. \ref{Transeq2}, as a function of frequency and temperature for a representative sample grown by the optical floating zone method with thickness $d$ = 22 $\mu$m. (b) Real part of the optical conductivity. $\sigma _1 (\omega, T)$, extracted from the transmission shown in (a) by numerical inversion of Eq. \ref{Transeq2}.}
\label{Fig1}
\end{figure}

\subsection{Optical Conductivity Of SmB$_6$}

Fig. \ref{Fig1}(a) displays the magnitude of the complex transmission, as defined in Eq. \ref{Transeq2}, as a function of temperature and frequency for a representative sample grown by the optical floating zone method with thickness $d$ = 22 $\mu$m.  Samples grown by the aluminum flux method display similar results but with poorer signal to noise due to smaller sample sizes and technical aspects of our measurement \cite{LauritaSmB6}.  Therefore, we focus here on the optical floating zone samples.  At the lowest temperatures, the largest transmission of $\approx$ 20\% is observed and then quickly decreases with increasing frequency.  The transmission also displays strong temperature dependence, decreasing with increasing temperature until becoming opaque in the THz range for temperatures T $\ge$ 30K for sample thicknesses $d$ $>$ 10 $\mu$m.   As we will discuss below these features are generally consistent with residual conductivity within a gap which is closing or filling in with increasing temperature.

As stated in the methods section above, the real and imaginary parts of the complex optical conductivity can be extracted from the complex transmission via numerical inversion of Eq. \ref{Transeq2}.  Fig. \ref{Fig1}(b) displays the real part of the optical conductivity, $\sigma _1(\omega , T)$, which was extracted from the transmission shown in Fig. \ref{Fig1}(a).  With some notable differences discussed previously \cite{LauritaSmB6}, the general frequency and temperature dependence of these data are in rough agreement with those of previously reported optical studies of SmB$_6$ \cite{Travaglini1984, Kimura1994, Nanba1993, Gorshunov1999}, although the exceptionally high resolution of our measurements provides new details.

\begin{figure*}
\includegraphics[width=2.0\columnwidth, keepaspectratio]{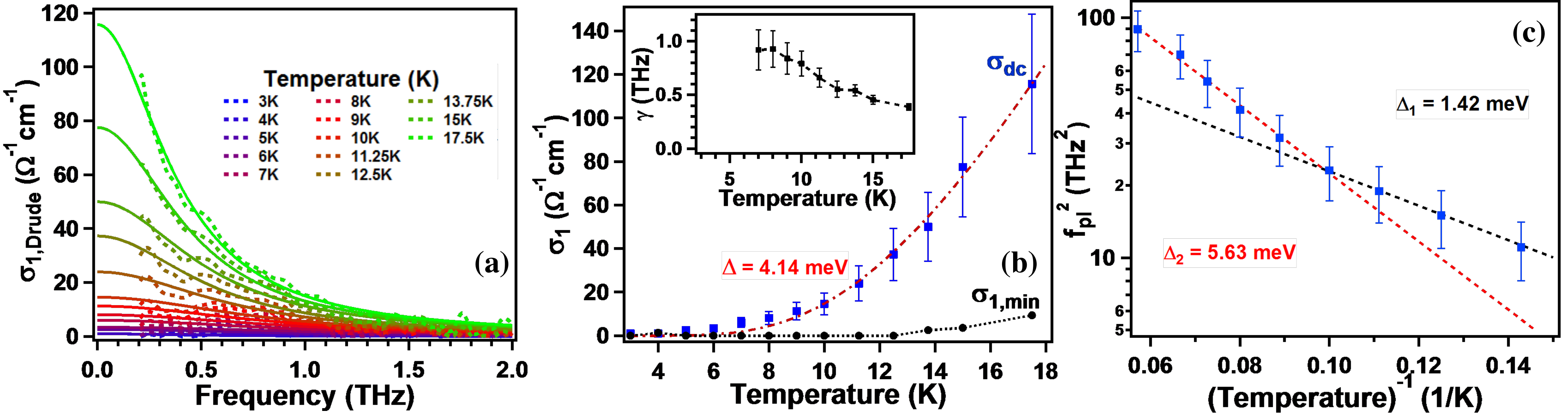}
\caption{Analysis of the free carrier response of SmB$_6$.  (a) Free carrier conductivity (dashed lines) with corresponding fits to the Drude model (solid lines) as a function of temperature.  (b) Temperature dependence of the dc conductivity (blue squares) with a corresponding exponential fit (red dashed line) from which an activation gap of 4.14 meV is extracted.  Also shown is the temperature dependence of the minimum conductivity (black circles).  Inset: Temperature dependence of the scattering rate.  (c) Arrhenius plot of the plasma frequency where two activation energy scales can be observed in distinct temperature regimes.  The black dashed line is a fit with $f_{pl}^2 $ $\propto$ $\exp{(-\Delta_1/k_BT)}$ while the red dashed line corresponds to a fit with $f_{pl}^2 $ $\propto$ $\exp{(-\Delta_2/2 k_BT)}$ such that a direct comparison to Ref \cite{Gorshunov1999} can be made.}
\label{Fig2}
\end{figure*}  

The conductivity displays a crossover from metallic to insulating behavior as a function of temperature.  At the highest measured temperatures a Drude-like response is observed as the optical conductivity is largest at the lowest frequencies and is a decreasing function of frequency thereafter.  The Drude-like response indicates the presence of free charge carriers in the conduction band.  As the temperature is reduced, the magnitude of the Drude response correspondingly decreases, disappearing at T $\approx$ 13K, at which point the conductivity is nearly frequency independent out to 2 THz.  At lower temperatures, T $<$ 13K, the conductivity becomes an increasing function of frequency, displaying approximately linear behavior below $\approx$ 1 THz.  This change in the functional dependence of the conductivity with frequency signifies a shift to a new dominant conduction mechanism consistent with a localized response within the gap.  Above 1 THz the conductivity saturates and displays little dependence with temperature or frequency.  

It is clear from the discussion above that the conductivity of SmB$_6$ in the THz range consists of at least two contributions, a Drude-like response from free charge carriers and additional conduction from the ``localized" response.  As mentioned above, we refer here to the non-Drude conductivity as being associated with a localized response as these states do not contribute to the dc transport.  However, if a neutral Fermi surface does exists within the gap of SmB$_6$ then the excitations themselves may be delocalized.  Thus ``localized" is simply terminology.  

Here we make the ansatz that the conductivity of the localized response is independent of temperature in our measurement range.  This assumption is supported by the weak temperature dependence of the conductivity at high frequencies, well above the scattering rate of the Drude contribution.  We can then model the total frequency and temperature dependent conductivity as:  
\begin{equation}
\sigma_{1,\text{Tot}}(\omega, T) = \sigma_{1,\text{Loc}}(\omega) + \sigma_{1,\text{Drude}}(\omega, T) + \sigma_{1,\text{min}}(T)
\label{CondModel}
\end{equation}
where $\sigma_{1,\text{Drude}}(\omega, T)$ is the free carrier response, $\sigma_\text{Loc}(\omega)$ is the temperature independent conductivity of the localized response, and $\sigma_\text{min}(T)$ is a frequency independent background conductivity which was found to be required to accurately fit the spectra above 12K.  It should be noted that this minimum conductivity has been previously included in fits of the optical conductivity of SmB$_6$ \cite{Gorshunov1999}, and interpreted in terms of a Mott minimum conductivity \cite{Mott1982}.  With the conductivity modeled in this fashion, we can now separate the free carrier and localized responses and analyze them independently.   

\subsection{Analysis Of The Free Carrier Response}

To isolate the free carrier response, we simply subtract the conductivity of the localized response from the total conductivity shown in Figure \ref{Fig1}(b).  We can identify the localized conductivity to be the total conductivity at the lowest measured temperature, T = 1.6K, at which point the free carrier conductivity is frozen out.  Subtracting the conductivity at T = 1.6K then isolates the free carrier response apart from the background $\sigma_{1,\text{min}}(T)$, which can be included in a fit and subtracted as well.  Figure \ref{Fig2}(a) displays the extracted free carrier conductivity (dashed lines) found by subtracting the 1.6K conductivity and $\sigma_{1, \text{min}}$ as described above.  As expected, the free carrier conductivity is largest at high temperatures and low frequencies.  As the temperature is reduced, thermal activation across the gap is frozen out and the corresponding free carrier conduction is significantly reduced.      

We can model the free carrier conductivity with the Drude model, in which case the frequency dependent ac conductivity is given by:
\begin{equation}
\sigma_{1,\text{Drude}}{(\omega)} = \frac{\sigma_\text{dc} \gamma^2}{(\gamma^2 + \omega^2)}
\label{Drude}
\end{equation} 
where $\sigma_\text{dc}$ = ${n e^2 \tau}/{m^*}$ = ${f_{pl}^2}/{2\gamma}$ is the dc conductivity, $\gamma$ = ${1}/{2 \pi \tau}$ is the scattering rate, $f_{pl}$ is the plasma frequency and $n$ and $m^*$ are the charge density and effective mass respectively.  Solid lines in Figure \ref{Fig2}(a) are fits of the spectra to Eq. \ref{Drude}.  One can see that the free carrier conductivity is well described by the Drude model, consistent with our overall conductivity model given in Eq. \ref{CondModel}.

We can then extract the temperature dependent dynamical properties of the free charge carriers from the Drude fits shown in Figure \ref{Fig2}(a).  Figure \ref{Fig2}(b) displays the temperature dependence of the extracted dc conductivity, which displays the expected activated behavior with temperature.  We can ascertain the activated energy scale by fitting the dc conductivity with the expression $\sigma_{\text{dc}}$ $\propto$ $\exp{({-\Delta}/{k_BT})}$ where $\Delta$ is the activation energy.  The red dashed line in Figure \ref{Fig2}(b) displays a fit of the dc conductivity with such an activated exponential.  Only data above 7K were used for the fit as the conductivity is most Drude-like in this temperature range.  From these fits we extract an energy scale of $\Delta$ = 4.14 meV, in excellent agreement with previous dc resistivity measurements \cite{Menth1969,Nickerson1971,Kim2013, Kim2014, Wolgast2013, Chen2015} and the $\Delta$ = 4.01 meV found in recent Corbino measurements, which have measured the bulk resistivity over 10 orders of magnitude \cite{Eo2017}.  Also shown in Figure \ref{Fig2}(b) is the temperature dependence of the minimum conductivity $\sigma_\text{1,min}(T)$, which is only finite above 12K, reaching a maximum value of 9.4 $\Omega ^{-1}$cm$^{-1}$ at 17.5K.  

The inset of Figure \ref{Fig2}(b) displays the scattering rate as a function of temperature.  The scattering rate is difficult to extract at low temperatures as the Drude conductivity quickly diminishes in magnitude.  However, the general trend of a slightly increasing scattering rate as the temperature is reduced is observed in the data down to 7K.  Below this temperature the error bars are too large to definitively assess the temperature dependence of the scattering rate and therefore this data is not included in the inset.  

Figure \ref{Fig2}(c) displays an Arrhenius plot the square of the plasma frequency, a quantity proportional to $n / m^*$.  Much like the dc conductivity, one expects $f_{pl}^2$ to possess activated behavior due to the temperature dependence of the charge density, although the implicit assumption here is that the effective mass is temperature independent over the temperature range of our measurement.  Fitting these data to similar activated expressions as the dc conductivity reveals two distinct regimes of activated temperature dependence with different energy gaps separated by an inflection point at 10 K.  Below 10K we find an activation gap of $\Delta_1$ = 1.42 meV while above 10K we find an activation gap of $\Delta_2$ = 5.63 meV.  These findings will be addressed in the discussion below.

\subsection{Analysis Of The Localized Response}

We identified the conductivity at the lowest measured temperature, 1.6K, to be the local response within the gap, as the Drude response is frozen out at these temperatures.  It was assumed that this conductivity is temperature independent at least up to 20K, as evidenced by the weak temperature dependence of the conductivity at high frequencies.  Here we analyze the conductivity of the localized response in more depth.

\begin{figure}
\includegraphics[width=1.0\columnwidth, keepaspectratio]{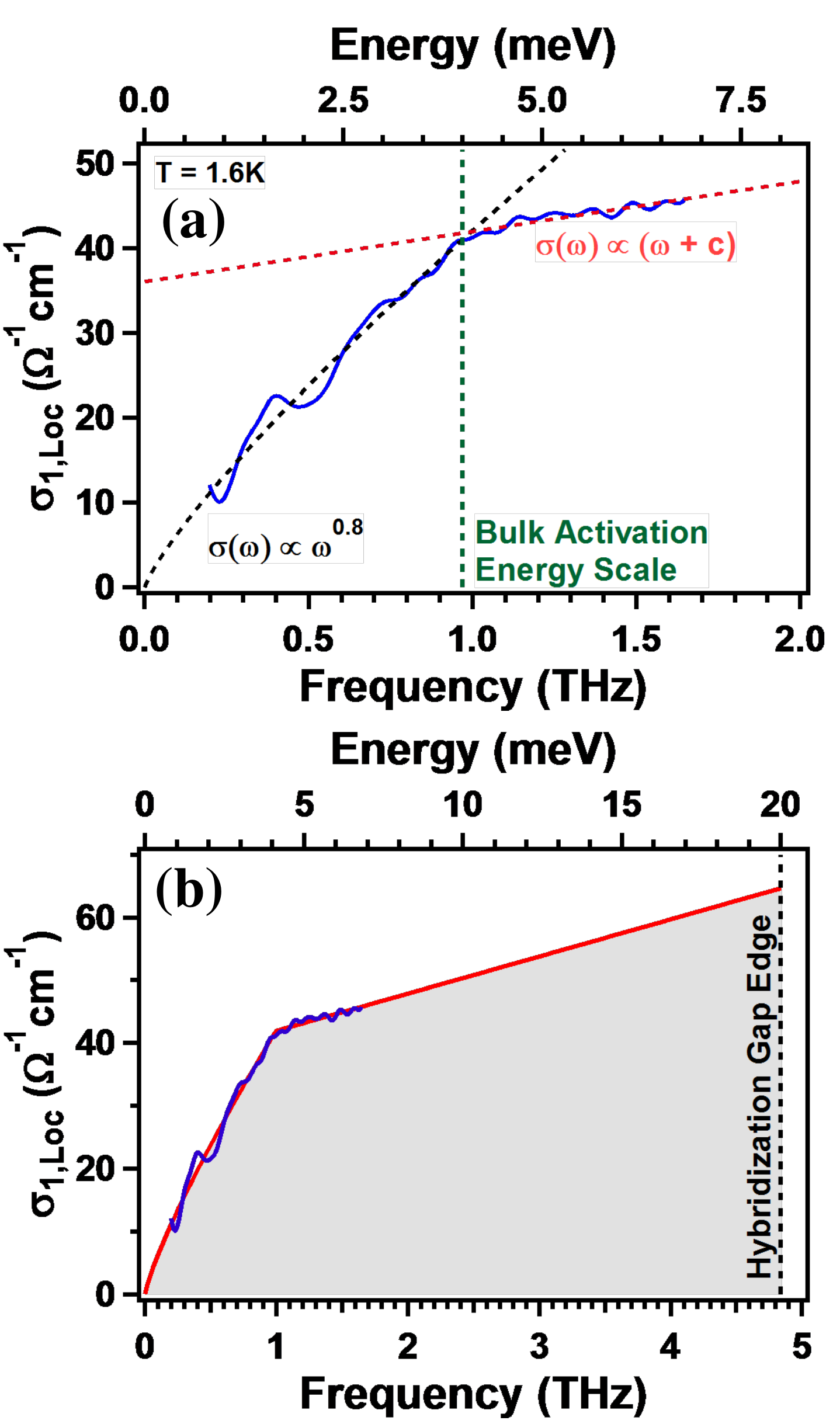}
\caption{(a) The real part of the conductivity of SmB$_6$ at 1.6K, identified here as the conductivity of the in-gap localized response. A shift in frequency dependence at the activated bulk energy scale is observed.  (b) The optical conductivity of the localized states extrapolated from zero frequency to the hybridization gap edge.  The gray shaded region represents a rough approximation of the spectral weight of this band from which the charge density or effective mass can be extracted, see text for details.}
\label{Fig3}
\end{figure}

Figure \ref{Fig3}(a) displays the localized conductivity as a function of frequency at T = 1.6K.  The conductivity can be separated into two regimes.  Below 1 THz, we find the conductivity is well described by a power law expression, $\sigma_{1, \text{loc}}(\omega)$ $\propto$ $\omega^n$ with an exponent $n$ $\approx$ 0.8.  Above 1 THz the conductivity shifts to a linear dependence with frequency, although fitting the conductivity over such a small frequency range is likely not definitive.  It should be noted that the energy scale of this crossover with frequency, $\approx$ 4 meV, is identical to the bulk activated energy scale observed in dc transport experiments.  We return to this observation in the discussion below.

We discussed the frequency dependence of this localized conductivity in some detail previously \cite{LauritaSmB6}. If such conduction stems entirely from impurity states then one might draw a parallel between SmB$_6$ and models of disordered Kondo insulators,  which have investigated the effects of substitutional doping of the $f$-ion sites \cite{Schlottmann1992,Riseborough2003,Mutou2001,Li1994}.  These models suggest that the Kondo gap is exceptionally sensitive to impurities as they break the translational invariance of the Kondo lattice and therefore destroy of the coherence of the heavy-fermion ground state.  For small impurity concentrations, these models predict an impurity band to form within the gap which scales with the square root of the impurity concentration.  At a critical impurity concentration, suggested to be as little as a 1\%, a percolation threshold between impurity states is reached, effectively closing the bulk gap.  The expected optical conductivity of these models depends on whether the impurity concentration is above or below this threshold.  Above the percolation threshold, the conductivity is expected to be a weak Drude peak for frequencies below the indirect gap, which is inconsistent with our observation of power law conductivity at low frequencies in SmB$_6$.  However, the conductivity below the percolation threshold, when the impurity states are presumably localized, has not yet been investigated to best of our knowledge.  

For a comparison to localized impurity states in disordered semiconductors, one may turn to localization driven insulators such as the disordered ``electron glass" Si:P \cite{Helgren2004}.  In such insulators, the expectation is that at the lowest temperatures ac conduction occurs between $resonant$ pairs of localized states.   Without interactions the ac conductivity is expected to follow Mott's famous $\omega^2$ law \cite{Mott1979}, which is clearly inconsistent with the data exhibited here.   With interactions included, but at frequency scales below that of the characteristic interaction energy between electron-hole pairs, the expectation is that the conductivity is quasi-linear with $\sigma_1 (\omega)$ = $e^4  D(\epsilon _F)^2 \xi^4 [\mathrm{ln}(2I_0/ \hbar \omega)   ]^3 \omega / \epsilon$ where $\xi$ is the localization length, $D(\epsilon _F)$ is the density of states at the Fermi level, and $I_0$ is the characteristic scale of tunneling between localized states that is expected to be bounded by the hybridization gap energy \cite{Shklovskii1981}.  Our ac conductivity data shown in Figure \ref{Fig3}(a) is roughly consistent with this model below 1 THz, with the caveat that the power law exponent is $n$ $\approx$ 0.8 instead of the expected linear dependence. However the $\omega^2$ dependence, expected at frequencies above the interaction energy scale of the system, is never recovered up to 2 THz.

However, whatever partial qualitative agreement exists between these data and that of localization driven insulators is overshadowed by the stark quantitative differences.  The in-gap ac conductivity of SmB$_6$ is $\approx$ 4 orders of magnitude larger than the impurity band conduction in Si:P (at say doped 39\% of the way towards the 3D metal-insulator transition) \cite{Helgren2004} and is essentially of the scale of the ac conduction in \textit{completely amorphous} Nb$_x$Si$_{1-x}$ alloys \cite{Helgren01a}.  Therefore, one must consider other possibilities.  Recent theories have proposed mechanisms by which one might attain a charge neutral Fermi surface within the Kondo gap \cite{Coleman1993,Baskaran2015,Onur2017,Chowdhury2017}.  Such neutral quasiparticles would be inert in dc transport experiments but may still couple to ac fields \cite{Ng07a}.  A separate theory claims that these in-gap localized states may originate from intrinsic electrons in SmB$_6$ that become self trapped through interactions with valence fluctuations \cite{Curnoe2000}.  Such theories have predicted functional forms for the optical conductivity in the context of their models but thus far none have predicted a power law behavior at low frequencies with exponents of n $\approx$ 1, as is observed in our measurements.

We can further characterize the origin of the localized carrier conduction by examining the spectral weight of the in-gap conductivity.  In an optics measurement, the spectral weight of a band can be related to the density of charge carriers $n$ and effective mass $m^*$ through the conductivity sum rule relation:
\begin{equation}
\int_0^W \sigma_{1}(\omega) \; d\omega = \frac{n e^2}{\pi m^*}
\label{fsum}
\end{equation}
where the upper limit of the integral is the unrenormalized electronic bandwidth \cite{Armitage2009,Mahan1990}.  Figure \ref{Fig3}(b) displays the localized conductivity with extrapolations of our fits of the data from zero frequency to 20 meV, roughly the expected value of the hybridization gap edge where it is known the conductivity rises by several orders of magnitude \cite{Gorshunov1999}.  The gray shaded area represents an estimate of the spectral weight of this band.  Using this area we may then make an estimate for the charge density of the localized response, although performing such an analysis requires knowledge of the effective mass.  Given the already crude nature of this approximation we assume that the effective mass is simply the free electron mass, although further discussion on this point will be made below.  Performing the calculation as described results in an extracted charge density of $n$ = 1.75 $\times$ 10$^{19}$ cm$^{-3}$ or $\approx$ 1 electron / 1000 unit cells.  This charge density will be analyzed in the context of impurities or a potentially exotic state in the discussion below.

\subsection{Discussion}

The analysis presented in this work is similar to that of a previous infrared conductivity study of SmB$_6$ \cite{Gorshunov1999}, which also separated the optical conductivity into free carrier and localized responses.  While aspects of the data presented here are in agreement with Ref. \cite{Gorshunov1999}, important differences exist, particularly in regards to the localized response.  Here we discuss how our results compare to previous studies and how our interpretations differ given recent developments in the understanding of SmB$_6$.

Our analysis of the free carrier response is in good qualitative agreement with the results of Ref. \cite{Gorshunov1999}, although quantitative differences exist.  For instance, the scattering rate reported in Ref. \cite{Gorshunov1999} is significantly smaller than that found in this work.  Gorshunov \textit{et al.} report a scattering rate of $\gamma$ $\approx$ 90 GHz below 15K, only roughly 20\% the $\gamma$ $\approx$ 0.5 THz observed here.  Additional differences are found in the temperature dependence of the scattering rate.  Gorshunov \textit{et al.} report a temperature independent scattering rate below 15K which was attributed to dominant impurity scattering.  Our extracted scattering rate displays a weak increasing behavior with reducing temperature from 20K to 6K.  It is unclear where this discrepancy originates.  It is possible that such an increase may be caused by slight temperature dependence of the localized conductivity, which we have assumed to be temperature independent in our model.  However, the validity of our model is evidenced by the fact that the extracted dc conductivity agrees with dc transport measurements performed on a sample from the same batch and displays the expected activated behavior with an energy scale in excellent agreement with dc transport experiments. 

Additional distinctions are found in the temperature dependence of the plasma frequency.  Both our work and Ref. \cite{Gorshunov1999} find the plasma frequency to display two distinct temperature regions of activated behavior with different activation energies.  However, the activated energies extracted in these two regimes are fairly different between our measurements.  Gorshunov \textit{et al.} report activation energies of 19 meV and 3 meV above and below 15K respectively, while our measurements indicate activation energies of 5.6 meV and 1.5 meV above and below 10K respectively.  One may be quick to associate such discrepancies to the difference in scattering rates but the weakly temperature dependent scattering rate is only a moderate modification compared to the exponential dependence of the dc conductivity.  Recalculating the plasma frequency with our extracted dc conductivity but with the scattering rate reported by Gorshunov \textit{et al.} results in activated energies of 12 meV and 2 meV, in better agreement but still only about 65\% the values reported in Ref. \cite{Gorshunov1999}. 

Our results regarding the localized response are in better quantitative agreement with Ref. \cite{Gorshunov1999}, although important differences still exist.  Gorshunov \textit{et al.} identify a ``bump" in the optical conductivity at 0.72 THz which was fit with a Drude-Lorentz oscillator and used to model the localized component of the optical conductivity.  The bump was attributed to an in-gap impurity band and thus all the localized conductivity was assumed to be related to impurities.  However, as seen in Figure \ref{Fig1}(b), we have not observed such a feature in any of our measurements of SmB$_6$ and speculate that this feature may have been an experimental artifact caused by standing wave resonances that occur in backwards wave oscillator setups as was used in Ref. \cite{Gorshunov1999}.  We therefore make no assumptions about the functional form of the localized conductivity with frequency in our analysis. It should be noted that despite this difference, Gorshunov \textit{et al.} find the localized conductivity to be only weakly temperature dependent below 20K, supporting our assumption that the localized contribution is temperature independent over the same range.

Such distinctions between our experimental results and that of Ref. \cite{Gorshunov1999} are important as they may alter the interpretation of the data.  Gorshunov \textit{et al.} attribute the different activation energy scales of 19 meV and 3 meV in the free carrier plasma frequency to be the hybridization gap and the energy gap between an impurity band and the bottom of the conduction band respectively.  Support for this interpretation was given by the bump in the localized conductivity at 3 meV, the same energy as the activation observed in the plasma frequency below 15 K.  However, as we have discussed, our experiments observe different activation behavior and no corresponding bump in the localized conductivity.  Moreover, Gorshunov \textit{et al.}'s bump was only a weak maximum in a large background of conduction, and it is not clear, even if such a band exists, why it would manifest in the dc data with an activated temperature dependence.  Given the results presented here, the activated energy scales in the plasma frequency may require reinterpretation.

Still, while we do not observe a maximum in the conductivity, we do observe a shift in its frequency dependence at the bulk activation energy scale as shown in Figure \ref{Fig3}(a).  The origin of this shift is unclear, although suggestions can be found in both impurity and neutral Fermi surface models.  For instance, Kondo impurity models predict that a finite impurity concentration smears out the peak in optical conductivity at the direct gap, leading to a tail of conduction which extends down to the indirect gap \cite{Riseborough2003}.  One may then interpret the 4 meV activation energy scale of the bulk as the indirect gap, with ac conduction below this energy scale stemming from localized impurity states which do not participate in the dc transport.  However, the same argument could be made in the case of a neutral Fermi surface, which, being charge neutral, would also be inert to dc transport experiments.

Additional information regarding the origin of the localized conductivity is provided by the spectral weight analysis shown in Figure \ref{Fig3}(b), from which we found an approximate charge density of $n$ $\approx$ 1 electron / 1000 unit cells.  It should be noted that such an analysis comes with large caveats.  In general, one must integrate up to frequencies of order the unrenormalized electronic bandwidth in order to infer the charge density or band mass from the conductivity sum rule.  Here the  integration was performed only to the hybridization gap edge, as the conductivity is known to rise by several orders of magnitude here \cite{Gorshunov1999}.  However, it is not obvious that the in-gap conduction must end at the hybridization gap edge.  Furthermore, our calculation was conducted under the assumption that the effective mass was that of the free electron mass, which likely is not the case as evidenced by quantum oscillations experiments \cite{Tan2015} which observe a mass of 0.18m$_e$ up to 10K.  Repeating the calculation with this effective mass lowers the extracted charge density to $n$ $\approx$ 1 electron / 5000 unit cells.  In contrast, if a ``heavy" mass is used then the charge density per unit cell can be significantly larger.  Thus, the uncertainty on this charge density is large.  Still, such an quasi-free electron analysis is enough to obtain an approximate understanding of the scale of the total number of charge carriers.  With these important considerations in mind, we now compare this value to the expectations of both impurity and neutral Fermi surface models.

Taken at face value this charge density is roughly consistent with what one might expect from impurities.  For instance, our extracted charge density is about 10 times larger than the charge density at the metal-insulator transition of Si:P \cite{Helgren2004}.  However, it is also about a factor of 10 less than the percolation threshold predicted by Kondo impurity models, which suggest that only a few percent impurity concentration is needed to effectively close the Kondo gap \cite{Riseborough2003,Mutou2001,Li1994}.  Given the large error bars on our extracted charge density, we can only remark that the extracted charge density is in general alignment with such theories.

A comparison can also be made to models of an in-gap neutral Fermi surface.  Before proceeding it should be noted that it is not obvious how the conductivity sum rule used in this fashion applies in the case of a neutral Fermi surface.  As we mentioned, performing the sum in Eq. \ref{fsum} to frequencies a few times the bandwidth in conventional metals allows one to extract the band mass.  What an analogous sum performed only over low frequencies in the case of a neutral Fermi surface tells us is, to the best of our knowledge, not yet known.  We hope that this analysis may inspire future theoretical investigations and we speculate that such a general statement about the sum rules may be more accessible than detailing the precise shape of the optical conductivity in such models.  

Still, comparisons can be made.  For instance, in the model of Refs. \cite{Chowdhury2017,Sodemann2017}, a neutral Fermi surface arises through fractionalization of $f$-holes into separate holon and spinon quasiparticles.  Strong binding between conduction $d$-electrons and the holons are proposed to form a charge neutral \textit{fermionic} composite exciton (FCE) which gives rise to a neutral Fermi surface in the appropriate limits.  In the context of this model the FCE density is found to be identical to the $d$-electron density, i.e. of order $n_\text{FCE}$ $\approx$ 1 / unit cell.  One may then insert this charge density into our sum rule analysis and instead ask what is the effective mass of these excitons.  Interpreting the result of our sum rule analysis in this fashion suggests an effective mass of $m^*$ $\approx$ 1000$m_e$.  How does this compare to the model of Ref. \cite{Chowdhury2017}? The effective mass of the FCE's may be inferred from the inverse of their hopping amplitude, $m^*_{\text{FCE}}$ $\propto$ $U_{df}$/$t_f t_d$, where $t_f$ and $t_d$ are the hopping amplitudes of the $f$-holes and $d$-electrons respectively and $U_{df}$ is the Coulomb repulsion between $f$ and $d$ electrons.  As the $f$ bands are nearly flat, $t_f$ is expected to be quite small, while the $U_{df}$ is expected to be a large energy scale of the system.  Thus, one expects a very large mass enhancement for the FCE's, consistent with the results of our sum rule analysis.

\section{Conclusion}

In this work, the optical conductivity of the Kondo insulator SmB$_6$ was shown to possess distinct regimes in which the conductivity is dominated by either free carrier or $localized$ contributions.  The free carrier response, dominant at temperatures above 10K, was analyzed in the context of the Drude model and found to be generally consistent with previous measurements, although quantitative differences were found and discussed.  The localized response, which becomes dominant at the lowest temperatures, was analyzed in the context of models of disordered insulators and recent developments which suggest a potentially neutral Fermi surface forms within the gap of SmB$_6$.  We found, through a conductivity sum rule analysis that the charge density or effective mass of this localized response is in reasonable agreement with models of both impurities and a neutral Fermi surface within the gap of SmB$_6$ in the appropriate limits.  Our results have shed further light on the mysterious in-gap conduction of SmB$_6$ and our hope is that these measurements will inspire future investigations into the properties of this remarkable material.

\section*{Acknowledgments}

Work at the Institute for Quantum Matter (IQM) was supported by the U.S. Department of Energy, Office of Basic Energy Sciences, Division of Materials Sciences and Engineering through Grant No. DE-FG02-08ER46544.  NJL acknowledges additional support through the ARCS Foundation Dillon Fellowship. TMM acknowledges support from The David and Lucile Packard Foundation.  We would like to thank D. Chowdhury, Y. S. Eo, P. Riseborough, and C. Varma for helpful conversations. 

\bibliography{SmB6Bib}

\end{document}